\documentclass[pra,twocolumn,superscriptaddress,showpacs]{revtex4}

\usepackage{graphicx}
\usepackage{amsfonts}
\usepackage{textcomp}

\newcommand{\beq}{\begin{equation}}
\newcommand{\eeq}{\end{equation}}

\begin{document}

\title{Time-resolved magneto-Raman study of carrier dynamics in low Landau levels of graphene}

\author{T. Kazimierczuk}
\author{A. Bogucki}
\author{T. Smole\'{n}ski}
\author{M. Goryca}

\address{Institute of Experimental Physics, Faculty of Physics,
University of Warsaw, ul. Pasteura 5, 02-093 Warsaw, Poland}

\author{C.~Faugeras} 
\address{Laboratoire National des Champs Magn\'etiques Intenses,
CNRS-UGA-UPS-INSA-EMFL, 25 rue des Martyrs, 38042 Grenoble, France}

\author{P. Machnikowski} 
\address{Institute of Physics, Wroc\l{}aw University of Technology, 50-370 Wroc\l{}aw, Poland}

\author{M. Potemski}

\address{Institute of Experimental Physics, Faculty of Physics,
University of Warsaw, ul. Pasteura 5, 02-093 Warsaw, Poland}
\address{Laboratoire National des Champs Magn\'etiques Intenses,
CNRS-UGA-UPS-INSA-EMFL, 25 rue des Martyrs, 38042 Grenoble, France}

\author{P. Kossacki} 

\address{Institute of Experimental Physics, Faculty of Physics,
University of Warsaw, ul. Pasteura 5, 02-093 Warsaw, Poland}

\email{Tomasz.Kazimierczuk@fuw.edu.pl}

\date{\today}

\begin{abstract}
We study the relaxation dynamics of the electron system in graphene flakes under Landau quantization regime
using a novel approach of time-resolved Raman scattering. The non-resonant character of the experiment allows
us to analyze the field dependence of the relaxation rate. Our results clearly evidence sharp increase in the
relaxation rate upon the resonance between the energy of the Landau transition and the G-band and shed new light
on relaxation mechanism of the Landau-quantized electrons in graphene beyond the previously studied Auger scattering.
\end{abstract}


\maketitle

\begin{figure}
\includegraphics[width=75mm]{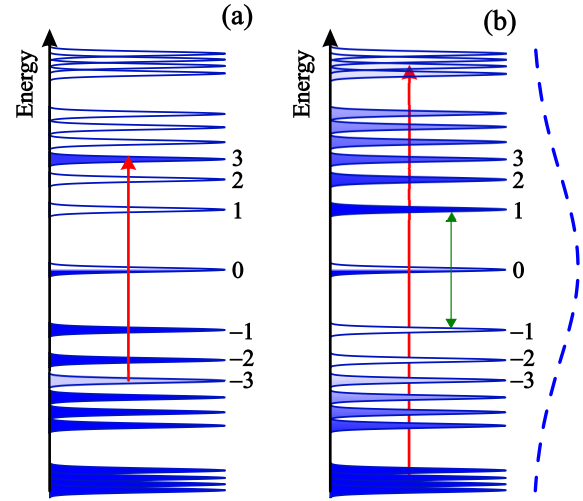}
\caption{Scheme presenting the spectrum of Landau Levels in graphene placed in external magnetic field. Color saturation denotes the relative population of LLs (a) directly after the pump pulse tuned to a transition between certain low-lying hole and electron LLs, (b) after a few hundred of fs following relatively high-energy (near infrared) laser pulse, during which the system reaches its quasi-equilibrium state due to fast Auger scattering.\label{fig:model_scheme}}
\end{figure}

Despite the whole rapidly expanding field of atomically thin semiconductors, graphene is still one of the most important systems with applications already being introduced. 
Carrier dynamics is one of a relevant issues in development of devices. Although they are directly related to the transport properties, optical tools are often needed to gain better insight into the carrier behavior. In particular, the ultra-fast dynamics of the carrier relaxation can be accessed using optical pump-probe techniques (for review see Ref. \onlinecite{Malic_book}). Although such an approach has been extensively exploited to study the basic problem of relaxation dynamics in graphene at zero magnetic field, an independent case of carrier relaxation between Landau Levels (LLs) emerging upon application of the magnetic field still remains to be relatively unexplored. In fact, there were only two time-resolved optical studies dealing with the carrier dynamics in Landau-quantized graphene~\cite{Plochocka_PRB_2009, Mittendorff_NatPhys_2015}. In both of these studies the pump and probe pulses were of the same energy. The pump pulse was utilized to initially populate a certain electronic LL, while the intensity of the probe was used to measure the dynamics of subsequent depletion of this LL (see Fig.~\ref{fig:model_scheme}(a)). Such a depletion was evidenced to be due to efficient Auger scattering regardless of the number of the excited LL, being either a high-energetic level ($n\sim100$) in the case of experiment exploiting near infrared Ti:sapphire laser~\cite{Plochocka_PRB_2009}, or a low-lying level ($n=0,1$) when the graphene was excited with a THz radiation produced by a free electron laser~\cite{Mittendorff_NatPhys_2015}.

\begin{figure}
\includegraphics{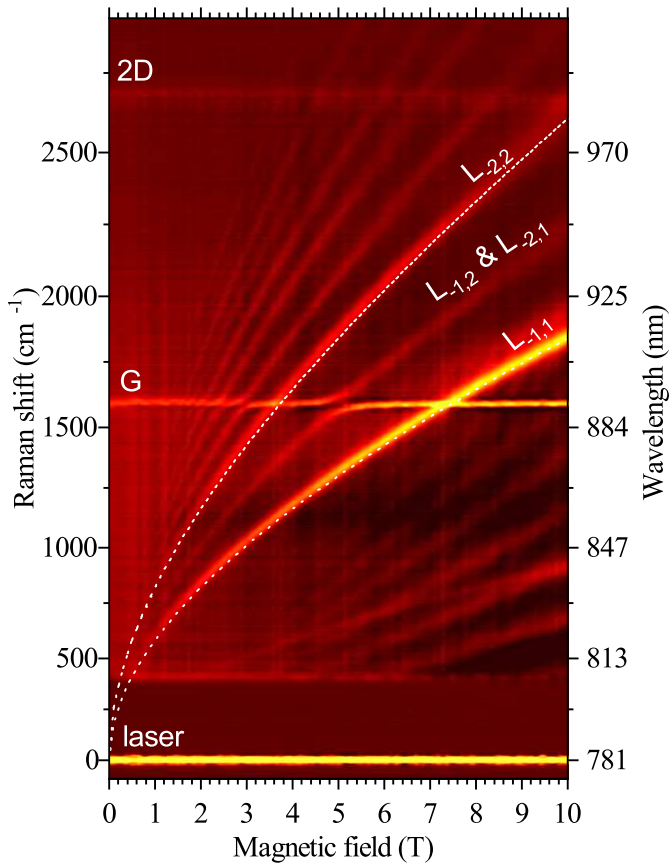}
\caption{Magnetic field dependence of the Raman scattering spectrum measured at $T=200$~K under excitation with CW Ti:sapphire laser at $\lambda=781$~nm. Each spectrum was corrected by subtracting 80\% of the zero-field spectrum in order to spotlight the field-induced changes. The white dashed lines mark two example Landau level transitions originating from graphene domain.  The signal below 400 cm$^{-1}$ is suppressed due to the long-pass filter placed in the detection path. \label{fig:sweep}}
\end{figure}

Here, we present a study of qualitatively different, slower carrier relaxation processes in Landau-quantized graphene, which take place \textit{after} the system reaches its quasi-equilibrium state due to fast Auger scattering (see Fig.~\ref{fig:model_scheme}(b)). Experimentally it is realized by combining pump-probe technique with monitoring the electronic transitions between LLs using Raman scattering spectroscopy. More specifically, a near infrared Ti:sapphire laser pulse is exploited to pump the carriers into some high-energy LL, from which they Auger-scatter occupying lower LLs, the population of which is finally measured based on the intensity of magneto-Raman peaks corresponding to electronic excitation between different LLs. Feasibility of such a technique was recently proven with respect to phonon transitions at zero magnetic field \cite{Yang_SREP_2016}. It has an advantage of high spatial resolution, inaccessible in previous experiment exploiting THz sources due to the difference in the wavelength scale. In particular, the presented approach allows to achieve sub-micrometer resolution in a standard microphotoluminescence setup. In such a regime, the intensity of various peaks in the Raman scattering spectrum provides direct access to carrier dynamics in a given LL even for very small graphene flakes, such as graphene domains on the surface of the natural graphite, which in turn exhibit much better optical properties as compared, e.g., to larger epitaxial graphene flakes \cite{Neugebauer_PRL_2009,Faugeras_PRL_2011}. Additional advantage of this system is its inherent neutrality, which makes the results clear from effects of residual background carriers, possibly present in the graphene system \cite{Sun_PRL_2010}.

In our work we studied graphene-like domains occurring on the surface of the natural graphite. The identification of such domains was performed upon application of strong magnetic field, which reveals qualitative differences in the LL structure between 2D graphene and 3D graphite \cite{Sadowski-PRL-2006, Faugeras_JRS_2018}. Figure \ref{fig:sweep}(a) presents the magnetic field evolution of the Raman scattering spectrum measured for one out of several investigated graphene domains. Apart from the well-known phonon-related resonances --- G-band $\approx 1590$ cm$^{-1}$ and 2D-band $\approx 2690$ cm$^{-1}$ --- the data features a series of field-dependent peaks corresponding to electronic excitations between different LLs. The optical selection rules allow transitions between $n$\textsuperscript{th} hole Landau level and $m$\textsuperscript{th} electron Landau level provided that $|n-m|\le1$ \cite{Kashuba-PRB-2009}. The energy position of such transitions in the spectrum is described by the square-root dependence characteristic for the Dirac dispersion\cite{Novoselov-Science-2004}: 
\begin{equation}
E_{-n,m}=\sqrt{2\hbar e B}v_F \left( \sqrt{\left|n\right|} + \sqrt{\left|m\right|} \right).
\end{equation}
The Fermi velocity extracted from the data shown in Fig. \ref{fig:sweep} yields $v_F=1.00\cdot 10^6$ m$/$s, which is comparable with the results reported in previous studies of such a system \cite{Faugeras_PRL_2011}. In the time-resolved experiments described in following sections, we focused mainly on the strongest Landau level transition from the 1\textsuperscript{st} hole level to the 1\textsuperscript{st} electron level denoted as $L_{-1,1}$.


The pronounced difference between the graphene and graphite magneto-Raman spectra allowed us also to spatially map the graphene domain. The lateral extension of the presented domain was found to be about 10~\textmu m~$\times$~25~\textmu m.



The core results presented in this work were obtained using two-color pump-probe Raman scattering spectroscopy technique in Landau quantization regime. The magnetic field needed for such experiments was applied by placing the sample inside a cryostat equipped with a superconductive magnet ($B=0 - 10$T) oriented in Faraday geometry. An aspheric lens, mounted on a piezo-positioners directly in front of the sample, allowed us to obtain spatial resolution of about 1~\textmu m. Such high resolution was important due to relatively small dimensions of the graphene domains as well as to achieve high pump laser fluence, which was needed for the time-resolved experiments. 

The probe beam used for Raman scattering spectroscopy was either femto- or picosecond Ti:sapphire laser at $\lambda_\mathrm{Ti:Sa}=775$~nm with 76 MHz repetition rate. The resulting Raman scattering signal was dispersed by a 30~cm-monochromator and recorded with a Si-based CCD camera. Dichroic filters in the excitation and the detection path were employed, respectively, to filter out the amplified spontaneous emission (ASE) and to remove the excess of the Rayleigh-scattered laser light.
Simultaneously, the sample was additionally excited with the second (pump) beam produced by optical parametrical oscillator (OPO) at $\lambda_\mathrm{OPO}=1200$~nm. The Raman signal induced by the OPO laser corresponds to IR range (e.g., $\lambda=1.48$~\textmu m for the G-band), and thus it was not detected in the experiment.

\begin{figure}
\includegraphics{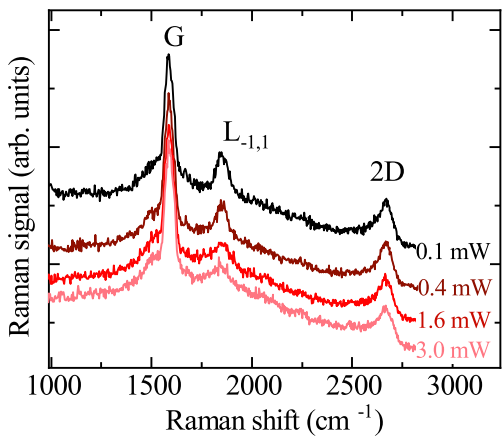}
\caption{A series of Raman scattering spectra measured using a pulsed laser of different intensity. Each spectrum was normalized using intensity of the G-band peak as the reference. Inset presents a dependence of the normalized intensity $L_{-1,1}$ on the excitation power. \label{fig:powerdep}}
\end{figure}

The OPO was pumped with the same Ti:sapphire laser that was used as a probe, which assured necessary synchronization between both laser pulse trains. The delay between pump and probe pulses was adjusted using a mechanical delay line. The overall temporal resolution of the experiments was limited by the duration of laser pulses. For the femtosecond configuration, used in most of the experiments, the FWHM of the pump pulses yielded 0.21~ps, while FWHM of probe pulses was equal to 0.44~ps. The main contribution to the latter value was the effect of the band-pass filter in the excitation path. Cross-correlation measurements of the pulses from both sources revealed no appreciable jitter, which would lead to reduction of the temporal resolution. In the picosecond configuration, the FWHM of the pump pulses yielded 3.7~ps, whereas the FWHM of the probe pulses was equal to 1.8~ps.


The LL population was studied by analysis of the intensity of the Raman $L_{n,m}$ peaks. Crucially, such a quantity is known to be proportional to the probability of the optical transitions between the involved levels, which become blocked for increasing occupancy of the LLs. As a result, the intensity of the Raman line starts to be quenched for sufficiently high carrier density, which in our case was controlled by changing the power of exciting laser. The invoked behavior is illustrated in Fig.~\ref{fig:powerdep}, which presents a set of Raman spectra measured at $B=10$~T using different intensities of the pulsed laser. Each of these spectra features two phonon peaks (G-band, 2D-band) as well as multitude of electronic peaks ($L_{-1,1}$, $L_{-1,2}$, \ldots). 
The intensity of phonon-related Raman peaks scales linearly with the excitation power. The underlying reason for such behavior is the high density of phonon states and their population can be affected significantly only using much stronger pump pulses \cite{Yang_SREP_2016}. In contrast, the electronic peaks exhibit aforementioned saturation behavior being due to filling the relevant electron or hole Landau levels, which, in turn, limits the density of states available for further Raman scattering \cite{Faugeras_JRS_2018,Kossacki_PRB_2012}. 

\begin{figure}
\includegraphics{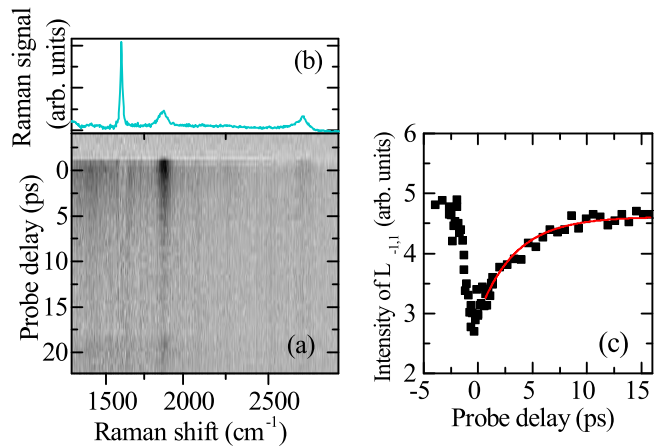}
\caption{(a) Spectrum of the changes induced by the pump beam in the Raman spectrum at $T=200$~K and $B=10$~T. The color reflects a difference between the measured Raman signal and the base signal at the same energy determined for the negative delay. The reference Raman spectrum on top of the map marks the position of various peaks in the spectrum. (b) Transient of the integrated intensity of the $L_{-1,1}$ peak. \label{fig:pumpprobe}}
\end{figure}

The same phenomenon was exploited in the pump-probe experiment: the strong pump pulse was utilized to initially populate the low Landau levels, while the intensity of the Raman peaks from the probe pulse was used as a measure of this population at later time. Example data measured in such an experiment are shown in Fig. \ref{fig:pumpprobe}. The power of the pump and probe beams was set to, respectively, 20 mW and 0.6 mW. The presented results clearly show that directly after the pump pulse the electronic Raman signal is weaker due to reduced density of states. Full spectrum of the changes [shown in Fig.~\ref{fig:pumpprobe}(a)] evidences that the pump indeed affects only $L_{n,m}$ peaks, while the phonon peaks (e.g., the G-band) do not exhibit noticeable variation upon arrival of the pump pulse. In agreement with previous studies~\cite{Mittendorff_NatPhys_2015}, the characteristic time scale of the pump-induced perturbation is in the range of a few picoseconds. Based on relatively fast (sub-picosecond) rise time of the signal, we attribute the decay dynamics directly to the relaxation rate of quasi-thermalized electronic system. The value of the relaxation time was extracted from the data by fitting exponential-decay profile to the measured transient. Example data shown in Fig. \ref{fig:pumpprobe}(b) yields decay time of $\tau= (3.4 \pm 1.0)$~ps. The employed experimental technique allowed us to follow the dynamics of the population of the Landau level continuously upon changes of the magnetic field, which was inaccessible in previous pump-probe experiments \cite{Mittendorff_NatPhys_2015}. 

\begin{figure}
\includegraphics[trim={0 3mm 0 0},clip]{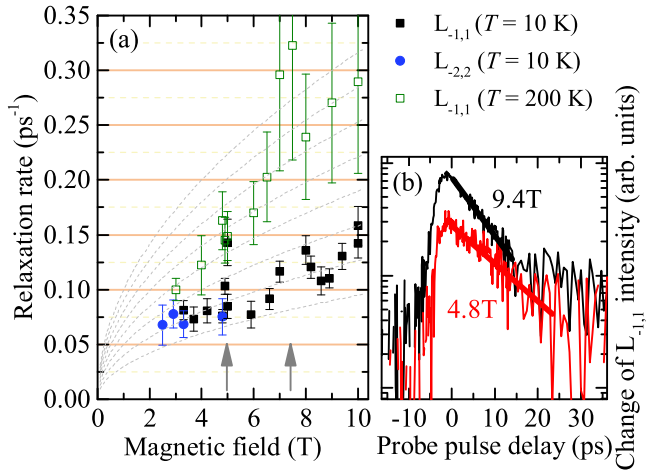}
\caption{(a) Electron relaxation times as a function of the magnetic field. A set of square-root functions are marked by dashed lines as a~guide to the eye. Two arrows indicate the resonant fields discussed in the text. (b) Pump-induced change in the signal for two magnetic fields demonstrating the difference in relaxation time. The straight lines represent the exponential-decay profiles fitted to experimental data. \label{fig:summary}}
\end{figure}

Two systematic data series are presented in Fig.~\ref{fig:summary}(a) for $T=200$~K and $T=10$~K. As seen, the data obtained for $L_{-1,1}$ and $L_{-2,2}$ transitions at lower temperature overlay each other, which is consistent with our assumption that the measured decay corresponds to a relaxation of the system remaining in a quasi-equilibrium with respect to Auger scattering. Importantly, we find that the rate of such a relaxation significantly increases around $B=5$~T and $B=7$~T (better visible for the lower temperature). This is the first observation of the theoretically predicted \cite{Wang-JPSJap-2013,Wendler-APL-2013} increase in the electron relaxation rate due to the resonance between the energy of the Landau levels and the $E_{2g}$ phonon. In particular, at $B=5$~T the resonance occurs for $L_{-2,1}$ and $L_{-1,2}$ transitions, while at $B=7$~T it is the $L_{-1,1}$ transition, which coincides with the energy of the invoked phonon. Surprisingly, in our data the resonance at 7~T is much broader than the one at 5~T. This finding remains in contrast with the previous theoretical predictions, according to which the resonant increase in the relaxation rate should occur rather for the non-symmetric transitions (e.g., $-1\to 2$) due to their strong mixing with the optical phonons \cite{Wang-JPSJap-2013,Wendler-APL-2013}. The reason for this disparity between the theory and the experiment is not clear at the moment.

The second observation on top of the resonant behavior discussed above is that, in general, the relaxation rate systematically increases with the magnetic field and it is much faster at higher temperature. Such finding might seem to be expected as several processes related to interaction with acoustic phonons exhibit similar increase of the rate with magnetic field and temperature. For example in many systems spin relaxation accelerates due to increase of the number of phonons accessible for higher relaxation energy \cite{standley-book}. Similarly the increase of the temperature results in the increase of population of acoustic phonons and higher probability of the relaxation. However the present case of the graphene is qualitatively different. The picosecond-time-scale relaxation of the LL occupation is related to cooling of the hot-electron system, which has to be mediated by electron-lattice energy transfer. The inter-landau level transition requires energy much higher than thermal energy even at moderate magnetic field (e.g., $1\to 0$ transition at $B=5T$ corresponds to 670K). Therefore the thermal population of active phonons is negligible and no significant variation should be observed for reasonable temperature range. Thus our experimental findings unequivocally show that the process cannot be explained by simple phonon-assisted relaxation between LLs and the relaxation is related to more complex processes involving low energy acoustic phonons and other higher energy excitations. The simplest mechanism is a two-phonon process, in which most of the energy is carried out by the optical phonon,
while acoustic phonons provide the required continuum. Such processes appear in the
second order in the carrier-phonon coupling or via optical phonon anharmonicity
\cite{Jacak2003}. 
The former relies on the electron-acoustic-phonon coupling.
An electron on the Landau level effectively interacts only with phonons of wavelengths not greater than
the magnetic length $l_{B}$, which restricts the available 
acoustic (in-plane) phonon energy to at most  $(v/v_{\mathrm{F}})E_{1}$, where $v$ is the
speed of sound. Since $v/v_{\mathrm{F}}\sim 10^{-2}$, the energy
conservation limits this two-phonon process to a very narrow range of inter-LL separations
around the optical phonon energy. Quantitative estimate is obtained by treating the two-phonon relaxation as an acoustic-phonon-mediated transition between $L_{-1,1}$ and electronic ground state with one optical phonon, which is enabled by an optical-phonon admixture to the LL states. The most resonant
optical phonon admixture to the $L_{-1,1}$ excitation is the phonon-assisted $L_{0,1}$ or
$L_{-1,0}$ state, 
since the most resonant single-phonon state on the electronic ground state (the G line in
the Raman spectrum in Fig.~\ref{fig:sweep}) is decoupled from the $L_{-1,1}$ electronic excitation, as
witnessed by the lack of resonant anti-crossing in the spectrum (which suggests that the
observed excitation is valley-symmetric \cite{Goerbig2007}). 
With only this admixture included and using the description of
carrier-phonon interaction in graphene \cite{Suzuura2008,Suzuura2009}, the maximum values
​​of the relaxation rate at 200~K, in the close vicinity of the resonant magnetic
field of $7.3$~T, are  comparable to those found in our measurements, while at 
10K the rates are a few orders of magnitude below the experimental values. In addition, the
theoretically predicted rate for this relaxation channel falls off exponentially and
decreases by many orders of magnitude already 1~T off resonance, in obvious contrast
with the measurements. 

The anharmonicity-induced relaxation channel yields rates smoothly varying with the
magnetic field, because the anharmonic decay of the zone-center optical phonon can involve
acoustic phonons with arbitrary, mutually opposite wave vectors. The resulting rate can be
estimated as the product of the optical phonon decay rate and the optical phonon admixture
to the LL. The former is determined, both experimentally \cite{Bonini2007} and
theoretically \cite{Kuhne2012} to be on the order of a few ps. The effective
optical-phonon-assisted coupling $V$ to other 
LLs, separated by energies on the order of the G-mode energy $E_{\mathrm{G}}$, can be
estimated as the typical with of a resonant anticrossing between the G line and the LL
excitations, which is on the order of a few meV. The admixture is then on the order of the
Huang-Rhys factor $(V/E_{\mathrm{G}})^{2}\sim 10^{-4}$, yielding the
anharmonicity-induced inter-LL relaxation in graphene ineffective. Quantitative
calculations indeed yield relaxation times on the order of 100~ns at magnetic fields
around 7~T.

We therefore conclude that the decay of the LL occupation has to be attributed to the overall cooling of the
hot-electron system via a more complex process, which might explain, in particular, the thermal dependence of the rate, which is compatible neither with the energies of the effectively cupled acoustic phonons nor with the inter-LL separation, nor with the optical phonon energy. 
As revealed by the data in Fig.~\ref{fig:summary}(a), the decay rate scales roughly as $\propto \sqrt{B}$ in the low temperature regime. Such a magnetic-field-dependence was theoretically predicted for broadening of LLs due to impurity induced dephasing~\cite{wendler-degruyter-2015}, which, in principle, could be responsible for the observed increase of the relaxation rate as long as it is accompanied by phonon scattering,
since the impurity dephasing alone is not expected to exhibit any temperature dependence~\cite{wendler-degruyter-2015}.
Another feature that is difficult to explain in terms of simple relaxation processes is
the pronounced change in the character of the magnetic-field dependence of the relaxation rate upon increasing the temperature, which is found to be almost linear at $T=200$~K. The detailed discussion of the physical reason for the observed features would require more in-depth knowledge about the nature of the involved relaxation processes, which demands further theoretical studies.
Possible mechanisms may involve relaxation between higher LLs, or combined phonon-Auger processes
involving those levels, accompanied by very fast redistribution of occupations between the
levels, consistent with the fs-time-scale rise of the Raman signal.

To conclude, our results demonstrate the feasibility of the time-resolved Raman scattering technique in studies of the Landau-quantized electrons. The non-resonant character of the Raman experiment enabled us to vary the magnetic field continuously, which is distinct advantage over previous approaches \cite{Mittendorff_NatPhys_2015}. The results of our study qualitatively confirm predictions regarding resonant increase in the electronic relaxation rate due to resonances with optical phonons. However, these results also reveal some deficiency of the existing theoretical description of the carrier dynamics in graphene at strong magnetic field. Further theoretical studies are needed to determine whether the detected discrepancy is related to the non-resonant character of our experiment or perhaps it is an indication of inadequacy of assumptions made in the existing models.

\begin{acknowledgments}
This work was supported by the Polish National Science Centre as research grant no DEC-2013/10/M/ST3/00791 and by the ATOMOPTO project carried out within the TEAM programme of the Foundation for Polish Science co-financed by the European Union under the European Regional Development Fund.
\end{acknowledgments}

\end{document}